\documentclass[reprint,showpacs,aps,nofootinbib,showkeys]{revtex4-1}
\usepackage{amssymb}
\usepackage{amsmath,amssymb,graphicx}
\usepackage{graphicx}
\usepackage{dcolumn}
\usepackage{bm}

\usepackage{amsbsy}
\usepackage{bm}
\usepackage{color}
\usepackage{epsfig}
\usepackage[colorlinks=true,linkcolor=blue,citecolor=red,urlcolor=blue,filecolor=blue]{hyperref}
\usepackage{slashed}
\usepackage{multirow}
\usepackage{slashed}

\def\beq{\begin{equation}}
\def\eeq{\end{equation}}
\def\bea{\begin{eqnarray}}
\def\eea{\end{eqnarray}}

\begin{document}

\bigskip

\vspace{2cm}
\title{Constraints on lepton number violating short-range interactions from $|\Delta L|=2$ processes}
\vskip 6ex

\author{N\'{e}stor Quintero}
\email{nquintero@fis.cinvestav.mx}
\affiliation{Departamento de F\'{i}sica, Centro de Investigaci\'{o}n y de Estudios Avanzados del IPN, Apartado Postal 14-740, 07000 M\'{e}xico D.F., M\'{e}xico}
\affiliation{Universidad Santiago de Cali, Facultad de Ciencias B\'{a}sicas, Campus Pampalinda, Calle
5 No. 62-00, C\'{o}digo Postal 760001, Santiago de Cali, Colombia}

\bigskip
\begin{abstract}	
In this work we study the short-range contributions that induce effective lepton number violating (LNV) interactions. We obtain a full set of constraints on the effective short-range couplings from a large variety of low-energy $|\Delta L|= 2$ processes of pseudoscalar mesons  $K, D, D_s, B$, and $\tau$-lepton. These constraints provide complementary and additional information to the one obtained from the neutrinoless double-$\beta$ ($0\nu\beta\beta$) decay.
As expected, the bounds on electron-electron short-range couplings  are the only ones that are strongly constrained by the $0\nu\beta\beta$ decay. Although weaker, LNV effective couplings with different flavours are not accessible to $0\nu\beta\beta$ decay and these can be probe by the $|\Delta L|= 2$ processes in consideration.
 \end{abstract}


\maketitle

\bigskip

\section{Introduction}  \label{Intro}

The observation of phenomena where the total lepton number $L$ is not conserved ($|\Delta L|=2$) remains as the best way to distinguish if neutrinos are Majorana fermions \cite{deGouvea:2013}. The experimental signal of such a lepton-number-violating (LNV) processes typically implies the production of same-sign di-lepton in the final state. Being forbidden within the Standard Model (SM), they would also be a clear indication of physics beyond the SM.

The neutrinoless double-$\beta$ ($0\nu\beta\beta$) decay has been regarded as the most appealing and sensitive test of such a LNV processes \cite{Rodejohann:2011,Gomez-Cadenas,Hirsch:2012,Vissani:2015,Deppisch:2015a}.
Observation of this nuclear decay would establish the existence of LNV processes, thus implying that neutrinos are Majorana particles \cite{Schechter:Valle,Takasugi-Nieves,Duerr:2011,Hirsch:2006}. 
Up to now, the $0\nu\beta\beta$ decay seems to be a rather elusive process and has not yet been observed experimentally. Currently, the best limits on their half-lives have been obtained from the nuclei $^{76}{\rm Ge}$ \cite{GERDA} and $^{136}{\rm Xe}$ \cite{EXO-200,KamLAND-Zen}. In the case when the exchange of a light massive Majorana neutrino (the so-called \textit{standard me\-cha\-nism} or \textit{mass mechanism} \cite{Rodejohann:2011,Gomez-Cadenas,Hirsch:2012,Vissani:2015}) is the dominant contribution to the $0\nu\beta\beta$ decay, the non-observation of these processes allow us to set constraints on the effective Majorana mass  at the sub-eV level ($\sim 10^{-1}$ eV) \cite{Rodejohann:2011,Gomez-Cadenas,Hirsch:2012,Vissani:2015}. 

Although the standard mechanism is considered as the most common interpretation, different new physics scenarios that generates LNV interactions can take place and therefore contribute to $0\nu\beta\beta$ decay. These are generically classified as the long-range \cite{Hirsch:1999,Hirsch:2016b} and short-range \cite{Hirsch:2001,Hirsch:2016} mechanisms. The standard interpretation belongs to the long-range one, while scenarios associated with heavy particle exchange are refered as \textit{non-standard me\-cha\-nisms} \cite{Rodejohann:2011,Hirsch:2012,Deppisch:2015,Hirsch:2013,Ge:2015} and they can be realized either through the long-range or short-range mechanisms \cite{Rodejohann:2011,Hirsch:2012,Deppisch:2015,Hirsch:2013,Ge:2015}.
Similarly to the case of the standard mechanism, the non-observation of $0\nu\beta\beta$ allows us to set model independent bounds on LNV effective couplings \cite{Hirsch:1999,Hirsch:2016b,Hirsch:2001,Hirsch:2016,Hirsch:2012,Deppisch:2015,Hirsch:2013,
Merle:2011}. 


Alternative $|\Delta L|=2$ processes to $0\nu\beta\beta$ have been proposed both at low and high energies as complementary evidence to prove the Majorana nature of neutrinos  (for a detailed list, see \cite{Atre:2009,Rodejohann:2011}), i.e. complementary test of the lepton number non-conservation \cite{Hirsch:2006}.
Among all these possibilities, the low energy studies of rare processes in $|\Delta L|= 2$ decays of pseudoscalar mesons and $\tau$-lepton have a\-ttrac\-ted a lot of attention \cite{Atre:2009,Kovalenko:2000,Ali:2001,Atre:2005,Kovalenko:2005,Helo:2011,Cvetic:2010,
Zhang:2011,Bao:2013,Wang:2014,Quintero:2016,Sinha:2016,Gribanov:2001,Quintero:2011,Quintero:2012b,Quintero:2013,Quintero:2016,Sinha:2016,Dong:2013,Yuan:2013,
Quintero:2012a,Dib:2012}, especially since these are accessible to different \textit{high-intensity frontier} experiments. According to their final state topology, they can be cla\-ssi\-fied as: 
\begin{itemize}
\item three-body channels \cite{Atre:2009,Kovalenko:2000,Ali:2001,Atre:2005,Kovalenko:2005,Helo:2011,Cvetic:2010,
Zhang:2011,Bao:2013,Wang:2014,Quintero:2016,Sinha:2016,Gribanov:2001} 
\begin{itemize}
\item $M^- \to M^{\prime +} \ell_\alpha^- \ell_\beta^-$,   
\item $\tau^- \to \ell_\alpha^+ M^{\prime -} M^{\prime\prime -}$ ,
\end{itemize}
\item four-body channels \cite{Quintero:2011,Quintero:2012b,Quintero:2013,Quintero:2016,Sinha:2016,Dong:2013,Yuan:2013,
Quintero:2012a,Dib:2012}
\begin{itemize}
\item $\bar{M}^{0} \to M^{\prime\prime +} M^{\prime +} \ell_\alpha^- \ell_\beta^-$ ,
\item $M^- \to M^{\prime\prime 0} M^{\prime +} \ell_\alpha^- \ell_\beta^-$  ,
\item $\tau^{-} \to M^{\prime +} \nu_{\tau} \ell_\alpha^- \ell_{\beta}^-$,
\end{itemize}
\end{itemize}

\noindent where $M \in \lbrace K, D, D_s, B, B_c \rbrace$ represents the decaying meson, $\alpha,\beta \in \lbrace e, \mu, \tau \rbrace$ are the leptonic flavours, and $M^{\prime}$ and $M^{\prime\prime}$ represent final hadronic states that are allowed by kinematics.

Experimentally, these $|\Delta L|= 2$ decays have been pursued for many years by different flavour facilities. No evidence has been seen so far and u\-pp\-er limits (UL) on their branching ratios have been reported by the Particle Data Group (PDG) and several experiments such NA48/2, BABAR, Belle, LHCb, and E791  \cite{PDG,NA48/2:2016,BABAR,BABAR:2014,LHCb:2012,LHCb:2013,LHCb:2014,Belle:2011,Belle:2013,E791}. At CERN, further improvements are expected by the NA62 kaon factory \cite{NA62} and the LHCb in the Run 2 and future upgrade Run 3 \cite{LHCbUpgrade}. In addition, the forthcoming Belle II experiment aims to get $\sim$ 40 times more data than the one accumulated by its predecessor Belle (as well as BABAR) \cite{BelleII}. All these efforts will increase the sensitivity on $|\Delta L|= 2$ signals by one or two orders of magnitude.


It is known that if the exchanged Majorana neutrino has a mass around $\sim$ 0.1 GeV to a few GeV, this might be produced on its mass shell and strongly enhance the $|\Delta L|= 2$ decays of pseudoscalar mesons and $\tau$-lepton \cite{Atre:2009,Kovalenko:2005,Helo:2011,Cvetic:2010,
Zhang:2011,Bao:2013,Wang:2014,Quintero:2016,Sinha:2016,Gribanov:2001,Quintero:2011,Quintero:2012b,Quintero:2013,Quintero:2016,Sinha:2016,Dong:2013,Yuan:2013,
Quintero:2012a,Dib:2012}. 
Although this GeV-scale sterile neutrino scenario is very interesting, it is worth exploring the possibility of other underlying short-range mechanism that could induce these LNV decays without involving Majorana neutrinos directly, just as it occurs in $0\nu\beta\beta$ decays.\footnote{For instance, the effect of a doubly-charged Higgs boson in the Higgs triplet model \cite{Quintero:2013b,Picciotto:1997}.} Since this latter can only probe LNV short-range couplings with electron flavour \cite{Hirsch:2001,Hirsch:2012,Deppisch:2015,Hirsch:2016}, we will provide bounds on the effective short-range couplings with the same ($\alpha=\beta$) or different ($\alpha\neq\beta$) leptonic flavour from the study of $|\Delta L|= 2$ decays: $M^{-} \to M^{\prime +} \ell_\alpha^- \ell_\beta^-$ and $\tau^{-} \to \ell_\alpha^+ M^- M^{\prime -}$, which are not accessible to $0\nu\beta\beta$ decay.

This work is organized as follows. In Sec. \ref{short_range} we briefly review the general aspects of the effective Lagrangian that describes short-range mechanisms. In Sec.  \ref{Delta L2} we study the constraints on the effective short-range couplings obtained from $|\Delta L|= 2$ decays of mesons and $\tau$-lepton. In Sec. \ref{IV} we discuss the similarities and differences of the present work in comparison with previous works. Our conclusions are left for Sec. \ref{Conclusion}.

\section{Short-range mechanisms} \label{short_range}

The short-range mechanisms refer to the effective interactions covering all processes mediated by heavy particles, in which no light neutrinos are exchanged \cite{Hirsch:2001,Hirsch:2012,Hirsch:2013,Deppisch:2015,Hirsch:2016}. 
The degrees of freedom of such a heavy particles are integrated out to get an effective 6-fermion vertex that induces LNV ($|\Delta L|= 2$) interactions.
Adopting the notation from Ref. \cite{Hirsch:2016}, the most general 6-fermion interaction $\bar{u}_i\bar{u}_jd_kd_n\ell_\alpha \ell_\beta$ (with arbitrary quark and
lepton flavours) is described by the short-range effective Lagrangian
\bea \label{Eff_Hamil}
\mathcal{L}_{\rm eff}^{\Delta L=2} &=& \dfrac{G_F^{2}}{2\Lambda} \sum_{i, X Y}  [C_i^{XY}]_{\alpha\beta} \mathcal{O}_i^{XY} ,
\eea

\noindent with $C_i$ the effective couplings (dimensionless) that generate LNV interactions and $\Lambda$ represents the mass scale dominant to the process in consideration (for instance, the proton mass in $0\nu\beta\beta$ decay). According to their Lorentz structure, the associated dimension-9 operators are classified as \cite{Hirsch:2016}
\bea 
\mathcal{O}_1^{XY} &=& 4 [\bar{u}_i P_{X} d_k] [\bar{u}_j P_{Y} d_n] j,  \\
\mathcal{O}_2^{XX} &=& 4 [\bar{u}_i \sigma^{\mu\nu}P_{X} d_k] [\bar{u}_j\sigma_{\mu\nu}P_{X} d_n] j,  \label{O2} \\
\mathcal{O}_3^{XY} &=& 4 [\bar{u}_i \gamma^\mu P_{X} d_k] [\bar{u}_j \gamma_\mu P_{Y} d_n] j,  \\
\mathcal{O}_4^{XY} &=& 4 [\bar{u}_i \gamma^\mu P_{X} d_k] [\bar{u}_j\sigma_{\mu\nu}P_{Y} d_n] j^\nu, \label{O4}  \\
\mathcal{O}_5^{XY} &=& 4 [\bar{u}_i\gamma^\mu P_{X} d_k] [\bar{u}_j P_{Y} d_n] j_\mu,
\eea

\noindent where $P_{X,Y}$ ($X, Y = L$ or $R$) are the chirality proyectors of the hadronic currents. The leptonic currents are defined as \cite{Hirsch:2016}
\bea \label{j_leptonica}
j &=& \bar{\ell}_\alpha (1 \mp \gamma_5) \ell^c_\beta, \\
j^\mu &=& \bar{\ell}_\alpha \gamma^\mu(1 \mp \gamma_5) \ell^c_\beta ,
\eea

\noindent with $\alpha,\beta \in \lbrace e, \mu, \tau \rbrace$. As it has been pointed out in \cite{Hirsch:2001,Hirsch:2012,Deppisch:2015,Hirsch:2016}, Eq. \eqref{Eff_Hamil} represents the most general, model independent, parametrization that can contributes not only to the $0\nu\beta\beta$ decay amplitude at tree level, but also to $|\Delta L|= 2$ processes involving any leptonic and hadronic state with second and/or third generation of quarks and leptons. 

At low energies, the parametrization \eqref{Eff_Hamil} is motivated from the nuclear physics point of view of $0\nu\beta\beta$ decay, a\-llo\-wing a finite set of combinations of six-fermion contact interactions (hadronic and leptonic currents) corresponding to a basic set of nuclear matrix elements \cite{Hirsch:2001,Hirsch:2012,Deppisch:2015,Hirsch:2016}. This approach is not unique and different effective o\-pe\-ra\-tor treatments can be considered \cite{deGouvea:2008,Babu:2001,Aparici:2012}. For ins\-tan\-ce, in a effective Lagrangian  approach \cite{Aparici:2012}, all virtual effects of a new physics scale ($\Lambda^{\prime}$) are proportional to $\lambda_{\alpha\beta}^{(9)} \mathcal{O}^{(9)}/\Lambda^{\prime 5}$, where $\lambda_{\alpha\beta}^{(9)}$ is the coefficient of the co\-rres\-ponding dimension-9 operator  $\mathcal{O}^{(9)}$ \cite{Aparici:2012}. So that, the  coefficients of  \eqref{Eff_Hamil} and the effective Lagrangian  approach are related by $[C_i]_{\alpha\beta} \sim 2  \lambda_{\alpha\beta}^{(9)} \Lambda/ \Lambda^{\prime 5}$ \cite{Aparici:2012}.

The $0\nu\beta\beta$ decay can only probe LNV couplings with $\alpha=\beta=e$, i.e. involving only the first fermion family \cite{Hirsch:2001,Hirsch:2012,Deppisch:2015,Hirsch:2016}. As we will present in the next section, alternative low-energy $|\Delta L|= 2$ decays allow us to set bounds on effective short-range couplings with the same or different leptonic flavour not accessible to $0\nu\beta\beta$ decay.

\section{$|\Delta L|= 2$ decays induced by Short-range interactions}  \label{Delta L2}

\begin{figure}[!t]
\centering
\includegraphics[scale=0.55]{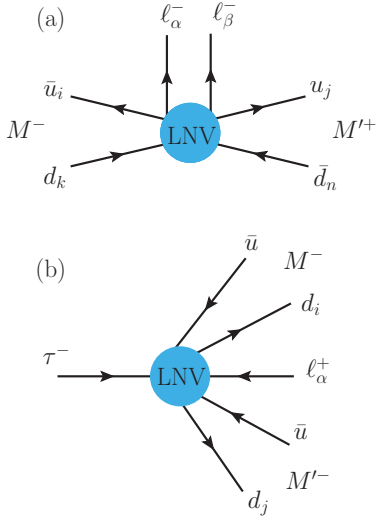}
\caption{\small $\Delta L = 2$ decays induced by short-range LNV operators: (a) $M^{-} \to M^{\prime +} \ell_\alpha^- \ell_\beta^-$ and (b) $\tau^{-} \to \ell_\alpha^+ M^- M^{\prime -}$. (See text for details).}
\label{Fig:DeltaL2} 
\end{figure}

Short-range interactions previously discussed can induce $\Delta L = 2$ processes to final or initial states containing leptons with the same or different flavour.  
In this section we consider the $|\Delta L|= 2$ decays: $M^{-} \to M^{\prime +} \ell_\alpha^- \ell_\beta^-$ and $\tau^{-} \to \ell_\alpha^+ M^- M^{\prime -}$, induced by short-range LNV (dimension-9) operators as shown correspondingly in Figs. \ref{Fig:DeltaL2}(a) and \ref{Fig:DeltaL2}(b). The mesons involved are generically denoted by $M^{(\prime)}  \in \lbrace \pi, K, D, D_s, B \rbrace$ and leptonic flavours by $\alpha,\beta \in \lbrace e, \mu \rbrace$. 
We will not deal with tensor currents, because of the antisymmetry of $\sigma_{\mu\nu}$ in \eqref{O2} and  \eqref{O4} the LNV tensor interactions are expected to be suppressed (vanishes to first order) \cite{deGouvea:2008}. So, we will focus only on operators $\mathcal{O}_i$ ($i=1,3,5$).

In order to obtain constraints on the effective short-range couplings $[C_i]_{\alpha\beta}$ from the non-observation of these $|\Delta L|= 2$ decays, it is phenomenologically reasonable to assume the dominance of only one short-range coupling, while the interference between different contributions is neglected \cite{Hirsch:2001,Hirsch:2012,Hirsch:2013,Deppisch:2015,Hirsch:2016}. As we will see, such a couplings will not depend on the quirality labels $X, Y$ and we will omit them since the beginning.

\subsection{$M^{-} \to M^{\prime +} \ell_\alpha^- \ell_\beta^-$} \label{Delta L2 meson}

By means of the short-range effective Lagrangian \eqref{Eff_Hamil}, the decay amplitude associated to $M^{-}(q) \to M^{\prime +}(q^{\prime}) \ell_\alpha^-(p) \ell_\beta^-(p^\prime)$ is obtained through the hadronization of the quark level $|\Delta L|= 2$ transition $\bar{u}_i d_k \to  u_j \bar{d}_n \ell_\alpha^{-} \ell_\beta^{-}$ [Fig. \ref{Fig:DeltaL2}(a)], as follows
\bea \label{3bodyM}
\mathcal{M}(M^{-} \to M^{\prime +} \ell_\alpha^{-} \ell_\beta^{-}) &=& \langle M^{\prime +}\ell_\alpha^{-}\ell_\beta^-| \mathcal{L}_{\rm eff}^{\Delta L=2} | M^{-}\rangle , \nonumber  \\
&=& \dfrac{G_F^2}{2 m_M} \sum_{i=1,3,5} [C_i]_{\alpha\beta} \ \mathcal{A}_i ,
\eea

\noindent where the mass scale $\Lambda = m_M$ and the matrix elements $\mathcal{A}_i$ are written as
\bea
\mathcal{A}_1 &=& F_{MM^\prime} \xi_M \xi_{M^\prime} \ [\bar{u}(p)(1 \mp \gamma_5) v(p^\prime)] , \label{A1} \\
\mathcal{A}_3 &=& F_{MM^\prime} (q\cdot q^\prime)[\bar{u}(p)(1 \mp\gamma_5) v(p^\prime)] , \label{A2} \\
\mathcal{A}_5 &=& F_{MM^\prime} \xi_M [\bar{u}(p)\slashed{q}^\prime (1 \mp \gamma_5) v(p^\prime)] \label{A3} , 
\eea

\noindent with $F_{M M^\prime} = V^{\rm CKM}_M V^{\rm CKM}_{M^\prime} f_M f_{M^\prime}$, where $V^{\rm CKM}_{M^{(\prime)}}$ and $f_{M^{(\prime)}}$ are the Cabibbo-Kobayashi-Maskawa (CKM) quark mixing matrix elements  and decay constants associated to the meson $M^-=\bar{u}_i d_k \ (M^{\prime +} = u_j \bar{d}_n)$. In the case of identical leptons ($\alpha=\beta$), it is necessary to add the antisymmetrized contribution  to \eqref{3bodyM}, which is obtained from the momentum exchange $p \leftrightarrows p^\prime$. Let us notice that we will take a phenomenological point of view and we will examine the structure of the LNV interactions in the quark mass basis rather than the weak basis, so that the CKM matrix elements appear explicitly in the above expressions.

To get expressions \eqref{A1}, \eqref{A2} and \eqref{A3}, we have used the hadronic parametrizations
\bea \label{Decayconstant}
\langle 0|\bar{d}_k \gamma^{\mu}\gamma_5 u_i| M \rangle &=& i f_{M} q^{\mu} , \\ 
\langle M^\prime |\bar{u}_j\gamma^{\mu}\gamma_5 d_n|0\rangle &=& -i f_{M^\prime} q^{\prime\mu} ,
\eea

\noindent for axial-vector current and
\bea   \label{Decayconstant_2}
\langle 0|\bar{d}_k \gamma_5 u_i| M \rangle &=& - i f_{M} \xi_{M} , \\ 
\langle M^\prime |\bar{u}_j \gamma_5 d_n|0\rangle &=& i f_{M^\prime} \xi_{M^\prime} ,
\eea

\noindent for the pseudoscalar one, with 
\bea  
\xi_{M} &=& \dfrac{m_M^2}{(m_{u_i}+ m_{d_k})}, \\ 
\xi_{M^\prime} &=& \dfrac{m_{M^\prime }^2}{(m_{u_j}+ m_{d_n})}.
\eea

\noindent From the above parametrizations, it is straightforward to see that effective short-range couplings $[C_i]_{\alpha\beta}$ will not depend on the quirality labels $X, Y$.

The decay witdh is given by
\bea 
\Gamma(&& M^{-} \to M^{\prime +} \ell_\alpha^{-} \ell_\beta^{-})  \nonumber \\ 
&& =\Big(1- \dfrac{1}{2}\delta_{\alpha\beta} \Big) \dfrac{	G_F^4}{128(2\pi)^{3} m_{M}^{5}}  \nonumber \\
&& \ \ \ \times   \Big[ \sum_{i=1,3,5} |C_i|^2_{\alpha\beta} \int_{s^-}^{s^+} ds \int_{t^-}^{t^+}  dt \ \big|\overline{\mathcal{A}_i} \big|^{2} \Big],
\eea

\noindent where $\big|\overline{\mathcal{A}_i} \big|^{2}$ are the squared matrix elements (spin-averaged), and $s \equiv m^2(\ell_\alpha^- \ell_\beta^-) = (p + p^\prime)^2$ and $t \equiv m^2(\ell_\beta^- M^{\prime +}) = (p^\prime + q^\prime)^2$ are kinematical (invariant masses) variables. Identical leptons in the final state are taken into account through the factor $(1-\delta_{\alpha\beta}/2)$. The integration limits are given by $s^{-} = (m_{\alpha}+m_{\beta})^2$, $s^{+} = (m_{M} - m_{M^\prime})^{2}$ and
\begin{eqnarray}
t^{\pm} &=& m_M^2 + m_{\alpha}^2 - \dfrac{1}{2 s} \Big[ (s + m_{M}^2 - m_{M^\prime}^2)(s + m_{\alpha}^2 - m_{\beta}^2) \nonumber\\
&& \mp \ \lambda(s,m_{\alpha}^2,m_{\beta}^2)^{1/2} \lambda(s,m_{M}^2,m_{M^\prime}^2)^{1/2} \ \Big],
\end{eqnarray}

\noindent with $\lambda(x,y,z)=x^{2}+y^{2}+z^{2}-2(xy-xz-yz)$ the usual kinematic function.

The non-observation of these $|\Delta L|= 2$ meson decays can be turned out into constraints on effective short-range interactions $|C_i|_{ee,\mu\mu,e\mu}$ (with $i=1,3,5$) as is shown in Table \ref{Constraints_1}. The CKM matrix elements, masses and decay constants of pseudoscalar mesons used in our calculations are listed in Table \ref{Inputs}.
As expected, in the case of the $ee$ couplings of short-range interactions, the experimental limits on $0\nu\beta\beta$ searches in $^{76}{\rm Ge}$ and $^{136}{\rm Xe}$ provide stronger bounds, typically of the order $(|C_1|_{ee}, |C_3|_{ee}, |C_5|_{ee}) \sim (10^{-7}, 10^{-8},10^{-7})$\footnote{The coeffecients $|C_i|_{ee}$ are equivalent to those usually denoted as $|\epsilon_i|$ \cite{Hirsch:2001,Deppisch:2015,Hirsch:2012}.} \cite{Hirsch:2001,Deppisch:2015,Hirsch:2012}, than those obtained from di-electron channels $M^{-} \to M^{\prime +} e^-e^-$. In the best case, this imply nearly eight (or higher) orders of magnitude above the sensitivity of $0\nu\beta\beta$ decays.

\begin{table}[t!]
\renewcommand{\arraystretch}{1.2}
\renewcommand{\arrayrulewidth}{0.7pt}
\caption{\small Constraints on effective short-range interactions $|C_i|_{\alpha\beta}$ (with $\alpha,\beta = e$ or $\mu$ and $i=1,3,5$) obtained from experimental UL on $M^{-} \to M^{\prime +} \ell_\alpha^- \ell_\beta^-$ \cite{PDG,NA48/2:2016,BABAR,BABAR:2014,LHCb:2012,LHCb:2013,LHCb:2014,Belle:2011}.}
\begin{tabular}{ccccc}
\hline\hline
Channel & Exp. UL & $|C_1|_{ee}$ & $|C_3|_{ee}$ & $|C_5|_{ee}$ \\
\hline
$K^- \to \pi^+ e^-e^-$ & $6.4 \times 10^{-10}$ & $3.3 \times 10^1$ & $2.3 \times 10^3$ & $2.9 \times 10^3$ \\
$D^- \to \pi^+ e^-e^-$ & $1.9 \times 10^{-6}$ & $4.3 \times 10^4$ & $3.2 \times 10^5$ & $1.2 \times 10^5$ \\
$D^- \to K^+ e^-e^-$ & $0.9 \times 10^{-6}$ & $1.8 \times 10^5$ & $8.9 \times 10^5$ & $3.8 \times 10^5$ \\
$D_s^- \to \pi^+ e^-e^-$ & $4.1 \times 10^{-6}$ & $1.7 \times 10^4$ & $1.2 \times 10^5$ & $4.5 \times 10^4$ \\
$D_s^- \to K^+ e^-e^-$ & $5.2 \times 10^{-6}$ & $1.1 \times 10^5$ & $5.3 \times 10^5$ & $2.3 \times 10^5$ \\
$B^- \to \pi^+ e^-e^-$ & $2.3 \times 10^{-8}$ & $6.0 \times 10^4$ & $1.4 \times 10^5$ & $6.2 \times 10^4$ \\
$B^- \to K^+ e^-e^-$ & $3.0 \times 10^{-8}$ & $3.0 \times 10^5$ & $6.0 \times 10^5$ & $2.6 \times 10^5$ \\
$B^- \to D^+ e^-e^-$ & $2.6 \times 10^{-6}$ & $3.6 \times 10^6$ & $5.2 \times 10^6$ & $2.8 \times 10^6$ \\
\hline\hline
Channel & Exp. UL & $|C_1|_{\mu\mu}$ & $|C_3|_{\mu\mu}$ & $|C_5|_{\mu\mu}$\\
\hline
$K^- \to \pi^+ \mu^-\mu^-$ & $8.6 \times 10^{-11}$ & $3.4 \times 10^0$ & $2.7 \times 10^2$ & $2.5 \times 10^2$ \\
$D^- \to \pi^+ \mu^-\mu^-$ & $2.2 \times 10^{-8}$ & $4.7 \times 10^3$ & $3.5 \times 10^4$ & $1.3 \times 10^4$ \\
$D^- \to K^+ \mu^-\mu^-$ & $1.0 \times 10^{-5}$ & $6.0 \times 10^5$ & $3.1 \times 10^6$ & $1.3 \times 10^6$ \\
$D_s^- \to \pi^+ \mu^-\mu^-$ & $1.2 \times 10^{-7}$ & $2.9 \times 10^3$ & $2.0 \times 10^4$ & $7.7 \times 10^3$ \\
$D_s^- \to K^+ \mu^-\mu^-$ & $1.3 \times 10^{-5}$ & $1.7 \times 10^5$ & $8.6 \times 10^5$ & $3.6 \times 10^5$ \\
$B^- \to \pi^+ \mu^-\mu^-$ & $1.3 \times 10^{-8}$ & $4.5 \times 10^4$ & $1.1 \times 10^5$ & $4.8 \times 10^4$ \\
$B^- \to K^+ \mu^-\mu^-$ & $5.4 \times 10^{-8}$ & $4.0 \times 10^5$ & $8.0 \times 10^5$ & $3.6 \times 10^5$ \\
$B^- \to D^+ \mu^-\mu^-$ & $6.9 \times 10^{-7}$ & $1.8 \times 10^6$ & $2.7 \times 10^6$ & $1.4 \times 10^6$ \\
$B^- \to D_s^+ \mu^-\mu^-$ & $5.8 \times 10^{-7}$ & $3.4 \times 10^5$ & $5.1 \times 10^5$ & $2.7 \times 10^5$ \\
\hline\hline
Channel & Exp. UL & $|C_1|_{e\mu}$ & $|C_3|_{e\mu}$ & $|C_5|_{e\mu}$ \\
\hline
$K^- \to \pi^+ e^-\mu^-$ & $5.5 \times 10^{-10}$ & $2.8 \times 10^1$ & $2.0 \times 10^4$ & $2.1 \times 10^3$ \\
$D^- \to \pi^+ e^-\mu^-$ & $2.0 \times 10^{-6}$ & $3.2 \times 10^4$ & $2.3 \times 10^5$ & $8.8 \times 10^4$ \\
$D^- \to K^+ e^-\mu^-$ & $1.9 \times 10^{-6}$ & $1.8 \times 10^5$ & $9.3 \times 10^5$ & $3.9 \times 10^5$ \\
$D_s^- \to \pi^+ e^-\mu^-$ & $8.4 \times 10^{-6}$ & $1.7 \times 10^4$ & $1.2 \times 10^5$ & $4.6 \times 10^4$ \\
$D_s^- \to K^+ e^-\mu^-$ & $6.1 \times 10^{-6}$ & $8.5 \times 10^4$ & $4.1 \times 10^5$ & $1.7 \times 10^5$ \\
$B^- \to \pi^+ e^-\mu^-$ & $1.3 \times 10^{-6}$ & $3.2 \times 10^5$ & $7.6 \times 10^5$ & $3.3 \times 10^5$ \\
$B^- \to K^+ e^-\mu^-$ & $2.0 \times 10^{-6}$ & $1.7 \times 10^6$ & $3.4 \times 10^6$ & $1.5 \times 10^6$ \\
$B^- \to D^+ e^-\mu^-$ & $1.1 \times 10^{-6}$ & $1.6 \times 10^6$ & $2.4 \times 10^6$ & $1.3 \times 10^6$ \\
\hline\hline
\end{tabular} \label{Constraints_1}
\end{table}

\begin{table}[t!]
\renewcommand{\arraystretch}{1.3}
\renewcommand{\arrayrulewidth}{0.7pt}
\caption{\small Numerical inputs: masses, CKM matrix elements, and decay constants.}
\begin{tabular}{cc}
\hline\hline
Meson masses (MeV) \cite{Rosner:2015} & Quark masses (GeV)  \cite{PDG} \\
\hline
$m_{\pi^\pm} = 0.1396$ & $m_u = 2.3$ MeV \\
$m_{K^\pm} = 0.4937$ & $m_d = 4.8$ MeV \\
$m_{D^\pm} = 1.8694$ & $m_s = 95$ MeV \\
$m_{D_s^\pm} = 1.9685$ & $m_c = 1.275$ GeV \\
$m_{B^\pm} = 5.279 $ & $m_b = 4.16$ GeV \\
\hline\hline
$f_P$ (MeV) \cite{Rosner:2015} & CKM elements \cite{PDG} \\
\hline
$f_\pi = 130.2$ & $|V_{ud}| = 0.97425$  \\
$f_K = 155.6$ & $|V_{us}| = 0.2252$  \\
$f_D = 211.9$ & $|V_{cd}| = 0.224$  \\
$f_{D_s} = 249.1$ & $|V_{cs}| = 0.966$ \\
$f_B = 187.0 $ & $|V_{ub}| = 3.67 \times 10^{-3}$  \\
\hline\hline
\end{tabular} \label{Inputs}
\end{table}


On the other hand, as it was previously mentioned, $0\nu\beta\beta$ decays do not allow us to put bounds on $\mu\mu$ and $e\mu$\footnote{This off-diagonal $|\Delta L|= 2$ transitions ($\alpha \neq \beta$) not only induced LNV processes but also induced lepton flavor violating (LFV) ones by one unit.} couplings of effective short-range interactions, since this can only probe the $ee$ ones. This is not the situation for the $|\Delta L|= 2$ channels $M^{-} \to M^{\prime +} \mu^-\mu^- (e^-\mu^-)$ that provide information on their corresponding short-range coefficients (see Table \ref{Constraints_1}). We observe that bounds are dictated by kinematics and CKM matrix elements involved. The most restrictive ones come from $K^- \to \pi^+ \mu^-\mu^- (e^-\mu^-)$. Although these bounds are too weak compared with those from $0\nu\beta\beta$ decay, in general, there is no fundamental theoretical reason for them to be of the same order since these test a different leptonic sector.
 
It  is worth mentioning that UL on the branching ratios of four-body decays $B^- \to D^0 \pi^+ \mu^-\mu^-$ \cite{LHCb:2012} and   $D^0 \to (\pi^- \pi^-, K^- \pi^-) \mu^+\mu^+$ \cite{,E791} can also be turned out into constraints on effective short-range interactions, which are expected to be similar to those reported in Table \ref{Constraints_1} and we have not included by simplicity.

\subsection{$\tau^{-} \to \ell_\alpha^+ M^- M^{\prime -}$} \label{Delta L2 tau}

Following a similar procedure as the previous section \ref{Delta L2 meson}, the decay amplitude of $\tau^{-}(p) \to \ell_\alpha^+(p^\prime) M^-(q) M^{\prime -}(q^\prime)$ is obtained through the hadronization of the quark level $|\Delta L|= 2$ transition $\tau^- \to  \ell_\alpha^{+} \bar{u}d_i \bar{u}d_j$ (Fig. \ref{Fig:DeltaL2}(b)) and it is written as
\bea \label{3bodytau}
\mathcal{M}(\tau^{-} \to \ell_\alpha^+ M^- M^{\prime -}) &=& \langle \ell_\alpha^+ M^- M^{\prime -}|\mathcal{L}^{\Delta L=2}_{\rm eff} | \tau^{-} \rangle , \nonumber  \\
&=& \dfrac{G_F^2}{2 m_\tau} \sum_{i=1,3,5} [C_i]_{\alpha\tau} \ \mathcal{T}_i ,
\eea

\noindent where the mass scale $\Lambda = m_\tau$ and the matrix elements $\mathcal{T}_i$ are defined by
\bea
\mathcal{T}_1 &=& F_{MM^\prime} \xi_M \xi_{M^\prime} \ [u(p)(1-\gamma_5) \bar{v}(p^\prime)] , \label{T1} \\
\mathcal{T}_3 &=& F_{MM^\prime} (q \cdot q^\prime)[u(p)(1-\gamma_5) \bar{v}(p^\prime)] , \label{T2} \\
\mathcal{T}_5 &=& F_{MM^\prime} \xi_M [u(p)\slashed{q}^\prime (1-\gamma_5) \bar{v}(p^\prime)]. \label{T3}  
\eea

\noindent For the case of identical mesons it is necessary to add the symmetrized contribution (interchanging $q \leftrightarrows q^\prime$) to \eqref{3bodytau}.  We have used Eqs. \eqref{Decayconstant} and \eqref{Decayconstant_2} to get the previous expressions.

Written in terms of kinematical variables $\tilde{s} = m^2(M^- M^{\prime -}) = (q + q^\prime)^2$ and $\tilde{t} = m^2(\ell_\alpha^+ M^{\prime -}) = (p^\prime + q^\prime)^2$, the decay rate is then given by
\bea 
\Gamma(\tau^{-} &&	\to \ell^+ M^- M^{\prime -}) \nonumber \\
&&= \Big(1- \dfrac{1}{2}\delta_{M M^\prime} \Big) \dfrac{	G_F^4}{256(2\pi)^{3} m_{\tau}^{5}}  \nonumber \\
&& \ \ \ \times \Big[ \sum_{i=1,3,5} |C_i|^2_{\alpha\tau} \int_{\tilde{s}^-}^{\tilde{s}^+} d\tilde{s} \int_{\tilde{t}^-}^{\tilde{t}^+}  d\tilde{t} \ \big|\overline{\mathcal{T}_i} \big|^{2} \Big]  ,
\eea

\noindent with $\big|\overline{\mathcal{T}_i} \big|^{2}$ the squared matrix elements (spin-averaged). The factor $(1-\delta_{M M^\prime}/2)$ accounts for identical mesons in the final state. In this case the integration limits are given by $\tilde{s}^{-} = (m_{M} + m_{M^\prime})^{2}$, $\tilde{s}^{+} = (m_{\tau} - m_\alpha)^2$, and
\begin{eqnarray}
\tilde{t}^{\pm} &=& m_\tau^2 + m_{M}^2 - \dfrac{1}{2 \tilde{s}} \Big[(\tilde{s} + m_{\tau}^2 - m_\alpha^2) (\tilde{s} + m_{M}^2 - m_{M^\prime}^2) \nonumber\\
&& \mp \ \lambda(\tilde{s},m_\tau^2,m_\alpha^2)^{1/2} \lambda(\tilde{s},m_{M}^2,m_{M^\prime}^2)^{1/2} \ \Big] .
\end{eqnarray}

Using the numerical inputs listed in Table \ref{Inputs}, in Table \ref{Constraints_2} we show the constraints that can be set  on effective short-range interactions $|C_i|_{e\tau,\mu\tau}$ ($i=1,3,5$), from the experimental UL on $|\Delta L|= 2$ decays of $\tau$-lepton. These off-diagonal short-range interactions also induce LFV interactions. In general, these bounds are of the same order to those obtained from $M^{-} \to M^{\prime +} \ell_\alpha^- \ell_\beta^-$ (see Table \ref{Constraints_1}) and too mild compared with those get from $0\nu\beta\beta$ decay. But, again, from the theoretical point of view it is not a priori clear that they have to be of the same order of the latter. In order to cover all the lepton flavours, it is important to point out that $\tau\tau$ coefficients (as well as $e\tau, \mu\tau$) might be explored in heavy meson decays $B_{(c)}^- \to \pi^+ \tau^-\tau^-$ \cite{Atre:2005,Atre:2009,Cvetic:2010,Sinha:2016}.

\begin{table}[t!]
\renewcommand{\arraystretch}{1.2}
\renewcommand{\arrayrulewidth}{0.7pt}
\caption{\small Constraints on effective short-range interactions $|C_i|_{\alpha\tau}$ (with $\alpha = e, \mu$ and $i=1,3,5$) obtained from experimental UL on  $\tau^{-} \to \ell_\alpha^+  M^{-} M^{\prime -}$ \cite{Belle:2013}.}
\begin{tabular}{ccccc}
\hline\hline
Channel & Exp. UL & $|C_1|_{e\tau}$ & $|C_3|_{e\tau}$ & $|C_5|_{e\tau}$ \\
\hline
$\tau^{-} \to e^+\pi^{-} \pi^{-}$ & $2.0 \times 10^{-8}$ & $3.4 \times 10^3$ & $5.0 \times 10^4$ & $8.8 \times 10^3$ \\
$\tau^{-} \to e^+\pi^{-} K^{-}$ & $3.2 \times 10^{-8}$ & $1.6 \times 10^4$ & $2.0 \times 10^5$ & $3.3 \times 10^4$ \\
$\tau^{-} \to e^+K^- K^-$ & $3.3 \times 10^{-8}$ & $1.4 \times 10^5$ & $1.5 \times 10^6$ & $3.6 \times 10^5$ \\
\hline\hline
Channel & Exp. UL & $|C_1|_{\mu\tau}$ & $|C_3|_{\mu\tau}$ & $|C_5|_{\mu\tau}$ \\
\hline
$\tau^{-} \to \mu^+\pi^{-} \pi^{-}$ & $3.9 \times 10^{-8}$ & $4.6 \times 10^3$ & $6.8 \times 10^4$ & $1.2 \times 10^4$ \\
$\tau^{-} \to \mu^+\pi^{-} K^{-}$ & $4.8 \times 10^{-8}$ & $2.0 \times 10^4$ & $2.5 \times 10^5$ & $4.1 \times 10^4$ \\
$\tau^{-} \to \mu^+K^- K^-$ & $4.7 \times 10^{-8}$ & $1.7 \times 10^5$ & $1.8 \times 10^6$ & $4.5 \times 10^5$ \\
\hline\hline
\end{tabular} \label{Constraints_2}
\end{table}

We close this section by mentioning that  the experimental non-observation of the $|\Delta L|= 2$ processes under study in this section (and previous one) can also be translated into lower limits on the scale of new physics responsible for the LNV interactions. By taking representative values for effective couplings of the order $\mathcal{O}(1)$, one can roughly estimate that LNV scale is of the order of $\mathcal{O}(5-50)$ GeV. This imply, in principle,  that such a low new physics scale would have already been seen at LEP, for instance, from rare $Z$-boson decays. However, if the search strategies were not sufficiently adequate, they could have escaped to the detection. If true, this open the possibility that they could still be there in this low energy scale and in that case, this will require a more dedicated search within reach of the high-intensity frontier experiments such a NA62, LHCb, Belle II and beam-dump (SHiP), rather than energy frontier.

\section{Comparison with similar works} \label{IV}

In the literature, most of the works attempting to establish constraints on the coefficients  of the dimension-9 effective operators have been dedicated to the $0\nu\beta\beta$ decay \cite{Cvetic:2010,deGouvea:2008,Babu:2001,Aparici:2012}, while there are only few works that have considered the bounds obtained from the $|\Delta L|= 2$ processes under study \cite{Cvetic:2010,deGouvea:2008}. In this section, we stress the similarities and differences of the present work compared with \cite{Cvetic:2010,deGouvea:2008}.

In Ref. \cite{deGouvea:2008} all effective LNV operators from di\-men\-si\-on-5  to dimension-11 has been studied. Based on dimensional arguments, these operators are analyzed in terms of an effective parameter $m^{\rm eff}_{\alpha\beta}$ (with $\alpha, \beta$ leptonic flavours), which is defined from the different classes of diagrams that contribute to the $|\Delta L|= 2$ decays in question. In the case of light Majorana neutrino exchange, this parameter is simply the effective Majorana neutrino mass. Within that treatment, the scale at which new physics appears can be estimated \cite{deGouvea:2008}.
In contrast, in this work we have paid attention to the dimension-9 LNV operator, particularly to the bounds on the respective e\-ffec\-tive couplings that can be set. In that sense, this work can be regarded as complementary to \cite{deGouvea:2008}.

On the other hand, concerning the Ref. \cite{Cvetic:2010}, the authors considered different LNV sources that incorporate short-range interactions, namely  left-right symmetric model (LRSM) and supersymmetry with R-parity violation (RPV) interactions, in $|\Delta L|= 2$ decays of mesons. For example, within LRSM the short-distance contributions arise from the exchange of heavy right-handed Majorana neutrinos and doubly-charged Higgs boson \cite{Cvetic:2010}. In both cases, one can identify that these contributions are generated by the dimension-9 effective operator $\mathcal{O}_3$, and the model independent bounds on the effective coeficients $|C_3|_{\alpha\beta}$ can be translated to bounds into the corresponding parameters of the LRSM. This also applies to the effect of a doubly-charged Higgs boson within the context of a Higgs triplet model \cite{Quintero:2013b,Picciotto:1997}.


\section{Conclusions} \label{Conclusion}

We have studied LNV ($|\Delta L|= 2$) interactions focusing on short-range contributions. In particular, we have set constraints on the effective short-range couplings $|C_i|_{\alpha\beta}$ (with the same $\alpha=\beta$ or different $\alpha \neq\beta$ leptonic flavour) from a large variety of low-energy $|\Delta L|= 2$ processes of mesons $M^{-} \to M^{\prime +} \ell_\alpha^- \ell_\beta^-$ and $\tau$-lepton $\tau^{-} \to \ell_\alpha^+ M^- M^{\prime -}$, which provide complementary and additional information to the one obtained from the $0\nu\beta\beta$ decay. The resulting bounds are summarized in Table \ref{Constraints_1} and \ref{Constraints_2}.
In the case of the coupling $|C_i|_{ee}$ ($i=1,3,5$), the experimental limits on $0\nu\beta\beta$ decays of nuclei $^{76}{\rm Ge}$ and $^{136}{\rm Xe}$ can provide stronger constraints than those obtained from di-electron channels $M^{-} \to M^{\prime +} e^-e^-$, nearly eight orders of magnitude above in the best case. While for the case of  short-range couplings $\alpha\beta = e\mu,\mu\mu,e\tau,\mu\tau$ (not accessible to $0\nu\beta\beta$ decay), 
we get that the most restrictive bounds are of the order $|C_i|_{\alpha\beta} \sim \mathcal{O}(1 -10^{2})$, which are still too weak compared with those get from $0\nu\beta\beta$ decay, showing that the electron couplings are the only ones that are strongly constrained. 
The significant progress that is expected by different high-intensity frontier experiments (NA62, LHCb, Belle II) will improved  by one or two orders of magnitude these bounds.


The obtained bounds on short-range couplings $|C_i|_{e\mu,\mu\mu}$ are generic and independent of models that incorporate LNV interactions. Those can be translated into particular realizations of non-standard mechanism \cite{Rodejohann:2011,Hirsch:2012,Deppisch:2015}, that leads to same-sign signals $e^-\mu^-$ and $\mu^-\mu^-$. 






\begin{acknowledgments}
The author thanks to Conacyt (M\'{e}xico) for financial support under project FOINS-296-2016 and Omar Miranda for useful suggestions.  He also acknowledges the support from Universidad Santiago de Cali.
\end{acknowledgments}



\end{document}